**On the early stages of precipitation during direct ageing of Alloy 718**

F. Theska[a], K. Nomoto[b], F. Godor[c], B. Oberwinkler[c], A. Stanojevic[c], S.P. Ringer[b], S. Primig[a,*]

[a] School of Materials Science & Engineering, UNSW Sydney, NSW, 2052, Australia

[b] The University of Sydney. Australian Centre for Microscopy & Microanalysis, and School of Aerospace, Mechanical and Mechatronic Engineering, The University of Sydney, NSW, 2006, Australia

[c] voestalpine BÖHLER Aerospace GmbH & Co KG, Kapfenberg, Austria

| | |
|---|---|
| * Corresponding author: | s.primig@unsw.edu.au |
| | tel.: +61 (0)2 9385 5284 |
| | fax: +61 (0)2 9385 6565 |
| | Room 346, E10 building, School of Materials Science & Engineering, UNSW Sydney, NSW, 2052, Australia |
| E-mail addresses: | f.theska@unsw.edu.au (F. Theska), |
| | keita.nomoto@sydney.edu.au (K. Nomoto), |
| | flora.godor@voestalpine.com (F. Godor), |
| | bernd.oberwinkler@voestalpine.com (B. Oberwinkler), |
| | aleksandar.stanojevic@voestalpine.com (A. Stanojevic), |
| | simon.ringer@sydney.edu.au (S.P. Ringer), |

**Abstract**

The Ni-based superalloy Alloy 718 is used in aircraft engines as high-pressure turbine discs and must endure challenging demands on high-temperature yield strength, creep-, and oxidation-resistance. Nanoscale γ'- and γ"-precipitates commonly found in duplet and triplet co-precipitate morphologies provide high-temperature strength under these harsh operating conditions. 'Direct ageing' of Alloy 718 is an attractive alternative manufacturing route known to increase the yield strength at 650 °C by at least +10%, by both retaining high dislocation densities and changing the nanoscale co-precipitate morphology. However, the detailed nucleation and growth mechanisms of the duplet and triplet co-precipitate morphologies of γ' and γ" during the direct ageing process remain unknown. We provide a correlative high-resolution microscopy approach using transmission electron microscopy, high-angle annular dark-field imaging, and atom probe microscopy to reveal the early stages of precipitation during direct ageing of Alloy 718. Quantitative stereological analyses of the γ'- and γ"-precipitate dispersions as well as their chemical compositions have allowed us to propose a model for the microstructural evolution. It is shown that fine γ'- and γ"-precipitates nucleate homogeneously and coarsen according to the Lifshitz-Slyozov-Wagner-theory. However, γ"-precipitates also nucleate heterogeneously on dislocations and experience accelerated growth due to Nb pipe diffusion. Moreover, the co-precipitation reactions are largely influenced by solute availability and the potential for enrichment of Nb and rejection of Al+Ti.

1. **Introduction**

Alloy 718 is a polycrystalline, precipitation-hardened Ni-based superalloy used as high-pressure turbine discs in aircraft engines [1,2]. More economical air travel is essential in the future as world annual air-traffic approximately doubles every two decades, where the key driver is the desire to cover longer distances with reduced travel time [3]. Therefore, increasing the aircraft engine efficiency for higher operational temperatures and workloads requires materials with improved high-temperature durability [4,5]. However, the development of novel alloys and/or advanced new thermo-mechanical processing of existing alloys are presently challenged to reach service cycles of >100,000 h at current operational temperatures of ~650 °C under the high mechanical service stresses [6,7].

The typical microstructure of Alloy 718 turbine discs consists of a γ-matrix with a grain size of ~ASTM 10, providing grain boundary and solid solution strengthening [8]. Grain boundaries are decorated with δ-phase platelets providing Zener-Smith pinning to improve creep resistance [9–11]. Nanoscale, long-range ordered γ'- ($Ni_3(Al,Ti)$ $L1_2$) and γ"- ($Ni_3(Nb)$ $D0_{22}$) precipitates contribute to the high-temperature strength via coherency and anti-phase boundary (APB) strengthening [12,13]. Duplet and triplet co-precipitate morphologies have been widely reported for Alloy 718 [14–16]. These morphologies are strongly influenced by the bulk Al, Ti, and Nb contents in Alloy 718 and derivatives. For instance, (Al+Ti)/Nb ratios ≥ 1 could promote γ' as primary precipitation event [15,17].

Conventional industrial manufacturing of Alloy 718 typically follows a routine of triple melting, homogenisation, forging, solution annealing, and age hardening [8]. However, so-called direct ageing (DA) omits the solution annealing step, and proceeds directly to age hardening after forging. This has been shown to provide yield strength increments of at least +10%, although the underlying microstructural processes have been revealed only recently [18–20].

While most studies have focused on conventionally processed Alloy 718, little is known about the detailed nucleation and growth processes of precipitates during the early stages of DA. Some of the current authors have shown that the triplet stacking sequence depends on the thermo-mechanical history, with γ"-precipitates commonly found in the centre of triplets in Alloy 718 processed via DA [20]. This is in contrast to conventional processing and indicates a change in the primary nucleation event. Sundararaman et al. [21] proposed simultaneous nucleation of γ' and γ", or initial γ"-nucleation based on co-precipitates identified via transmission electron microscopy (TEM) of conventionally processed Alloy 718. Alam et al. [22] used atom probe microscopy (APM) to demonstrate, that γ"-precipitation precedes the formation of γ'-precipitates in an Alloy 718 variant due to Nb-clustering. The nucleation of γ"-precipitates is also known to be sensitive to lattice imperfections in the γ-matrix, such as dislocations or stacking faults [21,23,24]. Compositional changes in clusters of Nb and non-stoichiometric γ"-precipitates suggest that the nuclei of the precipitating phases will possess different compositions [22,25]. The detailed nucleation events preceding the formation of the secondary or 'wing' precipitates that occur on a central or primary precipitate in the duplet and triplet co-precipitate structures also remain the subject of much discussion [14,16,26]. For instance, studies of Alloy 706 have revealed Ni-enrichments at the γ'/γ"-interfaces, suggesting that the isolated nucleation of the central precipitate and subsequent growth into the co-precipitate structures is a discontinuous process [26]. In contrast, APM studies on Alloy 718 by Miller et al. [14] suggest a continuous precipitation process such that the secondary phase nucleates directly at the γ'/γ-matrix, or the γ"/γ-matrix interface. Multi-phase field modelling has shown that secondary γ"-precipitates decorating primary γ'-precipitates may act as diffusion barriers and slow the coarsening of the central primary γ'-precipitate via hard impingement [16]. Thus, the aim of the present study is to clarify the sequence of events around the nucleation and growth of these co-precipitates, the role of dislocations, and the evolution of the chemical composition and co-precipitate morphology during DA of Alloy 718.

We provide a correlative microscopy approach for the quantitative characterisation of γ'- and γ"-precipitates during their early stages of precipitation. Hardness testing is applied to evaluate changes in the mechanical properties during ageing. Conventional TEM is used to resolve the precipitates on a global scale. High-resolution TEM (HRTEM) and high-angle annular dark-field (HAADF) imaging reveal γ'- and γ"-precipitates and their long-range ordering, as illustrated by a colorized HAADF image in Figure 1.

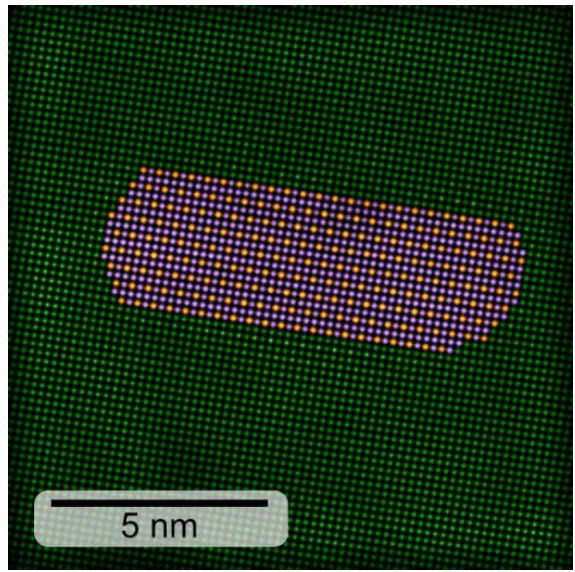

*Figure 1* Colourised HAADF image of a monolithic γ"-precipitate (Ni$_3$(Nb) purple, orange) in the γ-matrix (green).

Various APM techniques are employed to calibrate datasets and extract quantitative microstructural information such as precipitate volume fractions, radii, number density and morphology. These investigations have enabled us to provide a qualitative model for the microstructural evolution.

## 2. Material and Methods

Triple-melt (Vacuum induction melting VIM, electroslag re-melting ESR, and vacuum arc re-melting VAR), homogenised, forged and air-cooled Alloy 718 was provided by voestalpine BÖHLER Aerospace GmbH & Co KG, Austria. The chemical composition is provided in Table 1. After ingot breakdown and radial forging, the material was industrially forged from a billet of $h_{Billet}$ = 180 mm and $r_{Billet}$ = 127 mm into a pancake of $h_{Pancake}$ = 72 mm and $r_{Pancake}$ = 200 mm.

*Table 1 Chemical composition of Alloy 718.*

| Element | Ni | Fe | Cr | Nb | Mo | Ti | Al | C | B |
|---|---|---|---|---|---|---|---|---|---|
| wt.% | Balance | 18.5 | 18.0 | 5.3 | 3.0 | 0.95 | 0.50 | 0.02 | 0.003 |
| at.% | | 19.2 | 20.1 | 3.3 | 1.8 | 1.15 | 1.08 | 0.01 | 0.002 |

The thermo-mechanical processing followed industrial standards and the corresponding time-temperature-deformation routine is provided in Figure 2. Cuboidal samples in the dimensions 7x7x20 mm$^3$ were cut at the radial position $r_{Pancake}$ ~142 mm from the pancake centre. The ageing treatments were carried out in a Labec alumina tube furnace under Ar-atmosphere with a flow rate of 0.2 l/min and were followed by water quenching. The following DA conditions were acquired: 0 h (as-forged), 0.2 h, 0.6 h, 1.2 h, 2.3 h, 4.5 h, 9.1 h, 10.5 h, 11.9 h, 15.8 h, and 19.1 h (complete) DA. Ageing I and II were at 720 °C and 620 °C, respectively. The transition was furnace cooling with -0.5 °C/min.

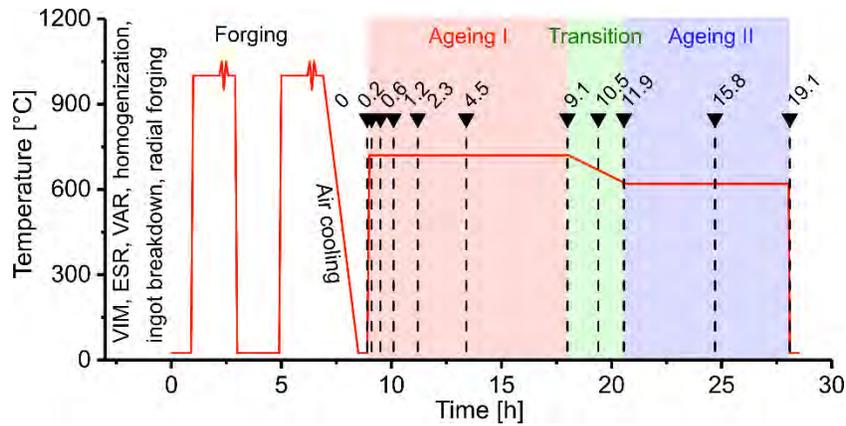

*Figure 2 Time-temperature-deformation schedule for DA of Alloy 718. Each marker represents one sample directly aged for the indicated time in hours.*

Vickers hardness HV10 testing was carried out using a STRUERS Duramin-300 according to the ASTM standard E92-82. Five indents per sample were measured to obtain the average hardness and standard deviation.

TEM foils were prepared using standard techniques [27]. Bright field (BF) and dark field (DF) images, and diffraction patterns were recorded with a JEOL JEM2100 TEM at 200 kV acceleration voltage for grains in <100> orientation. HRTEM images were recorded with a ThermoFisher Themis Z double aberration-corrected TEM at 300 kV acceleration voltage. Energy-dispersive X-ray spectroscopy (EDXS) mapping and HAADF were carried out for grains in <100> orientations. TEM and HRTEM images were post-processed using Fast-Fourier-Transform (FFT) tools and the TEM suite for FIJI ImageJ [28,29].

APM samples were prepared by cutting rods in the dimensions 0.5x0.5x20 mm$^3$ and electropolished following standard techniques [30]. Samples were stored in the load-lock chamber of the atom probe immediately post-preparation to minimise oxidation. A CAMECA LEAP 4000X Si straight flight path atom probe was used for voltage-mode and laser-assisted experiments. Our experiments were conducted at a temperature of 50 K, pulse rate of 200 kHz and a target evaporation rate of 1.0%. The voltage pulse rate was 20%, and the laser energy was 50 pJ. The CAMECA IVAS 3.6.6 software was used for data analysis. Voltage-mode APM datasets were calibrated using atom probe crystallography routines [31], and laser-assisted mode datasets were calibrated based on the precipitate morphology obtained from voltage-mode [32]. Full-size datasets approximated the tip volume as truncated cone to determine the precipitate volume fractions, radii, number density, morphology and chemical composition. The datasets were cropped to identical volumes for improved visualisation, determination of the bulk chemical composition, and the interface-method [32]. Figures 7 and 8 are projections of datasets which were cropped to cylinders of 30x30x50 nm$^3$ in voltage-mode, or 50x50x300 nm$^3$ in laser-assisted mode. Therefore, all of the APM data shown in this study are either 30 nm (voltage-mode) or 50 nm (laser-assisted) thick. Details to the enhanced iso-surface and interface methods are provided elsewhere [32]. Here, one-dimensional concentration profiles were extracted from 5 to 10 γ'-γ" duplet precipitates in perpendicular direction to the γ'/γ" interface, and four slices of 5 nm thickness were extracted per cylindrical dataset, respectively. The presented precipitate statistics were extracted from two voltage- and laser-mode APM datasets, respectively.

## 3. Results

### 3.1. Hardness testing

Figure 3 shows a semi-logarithmic plot of the hardness as a function of the ageing time. The as-forged condition is shown as 0 h DA with 350 ± 2 HV10. After 19.1 h (complete DA), the hardness increases to 480 ± 2 HV10. The other conditions follow a roughly exponential growth function as indicated by the dashed light red line. The 2.3 h to 9 h DA conditions somewhat deviate from this trend and approach a 'saturation point' before the transition. The transition describes an increase from 459 ± 2 HV10 to 473 ± 2 HV10. The four conditions indicated by black arrows were selected for TEM and APM analysis.

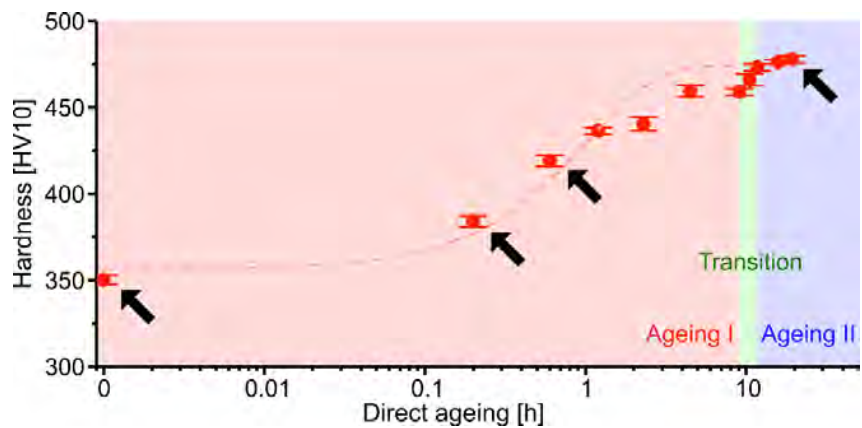

*Figure 3 HV10 evolution for DA of Alloy 718 including Ageing I, Transition, and Ageing II. Black arrows highlight the four conditions selected for TEM and APM analysis.*

## 3.2. Transmission electron microscopy

Figure 4 provides the <100> BF and <100> g(200)γ" DF TEM images with corresponding diffraction patterns as insets. The as-forged condition reveals mottled contrast in both the <100> BF and <100> g(200)γ" DF modes (Figures 4 a-b). Fine and coarse precipitates are highlighted by small blue and large red arrows, respectively. The reflections in the diffraction pattern correspond to the FCC γ-matrix, and both the $L1_2$ γ'- and $D0_{22}$ γ"-precipitate superlattices, which are marked by yellow circles. The 0.2 h DA condition in Figures 4 c-d exhibits a similar mottled contrast in both <100> BF and <100> g(200)γ" DF modes. However, coarse particles are more pronounced and decorate dislocation networks as planar arrays. The diffraction pattern contains both FCC and weaker superlattice reflections. In the 0.6 h DA condition (Figures 4 e-f), the γ'- and γ"-precipitates appear more clearly in both <100> BF and <100> g(200)γ" DF modes. Fine precipitates are densely arranged in the γ-matrix and planar arrays of coarse particles decorate the dislocation network. The diffraction pattern contains both FCC and ellipsoid γ'- and γ"-precipitate superlattice reflections. The complete DA condition in Figures 4 g-h reveals γ'- and γ"-precipitates in duplet and triplet structures. The reflections in the diffraction pattern resemble both FCC and γ'- and γ"-precipitate superlattices. Meaningful information considering particle radii and volume fraction could not be extracted from these images due to insufficient contrast.

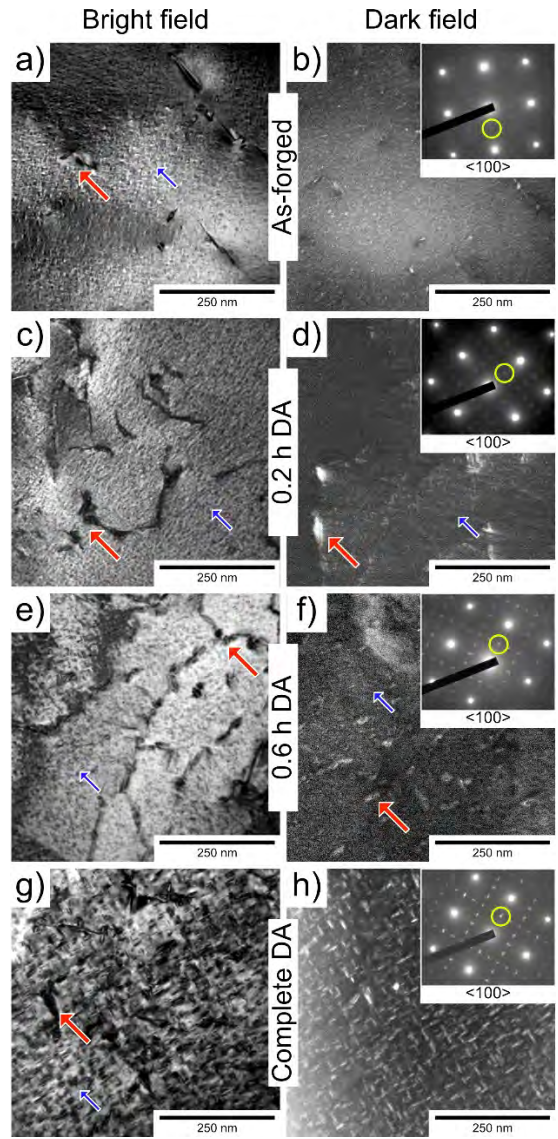

*Figure 4 TEM <100> BF and <100> g(200)γ" DF images with diffraction pattern insets. Fine and coarse γ'- and γ"-precipitates provide superlattice reflections in the diffraction patterns.*

STEM and HAADF images provide more detailed insights into the γ'- and γ"-precipitate nanostructure. A (purposely) defocused HAADF image of the as-forged condition shows fine and coarse precipitates in Figure 5 a. The correlating STEM image in Figure 5 b exhibits periodically arranged bright atom columns within fine precipitates (yellow outline). Alternating maxima in the intensity plot correlate to γ"-precipitates. The maxima in the FFT inset are associated to the FCC γ-matrix, and γ'- and γ"-precipitate superlattices (red arrows). Figures S1 a-e provide evidence of dislocations in the proximity of coarse γ"-precipitates.

Figure 5 c provides a (purposely) defocused HAADF image of the 0.6 h DA condition. Coarse precipitates appear bright, and fine precipitates exhibit duplet and triplet morphologies. The corresponding STEM image in Figure 5 d shows a facetted γ"-precipitate (yellow outline) and the intensity plot reveals the superlattice structure. The FFT-image corresponds to the FCC γ-matrix and γ'- and γ"-precipitate superlattices. Figures S1 f-j also show dislocations at coarse γ"-precipitates after 0.6 h.

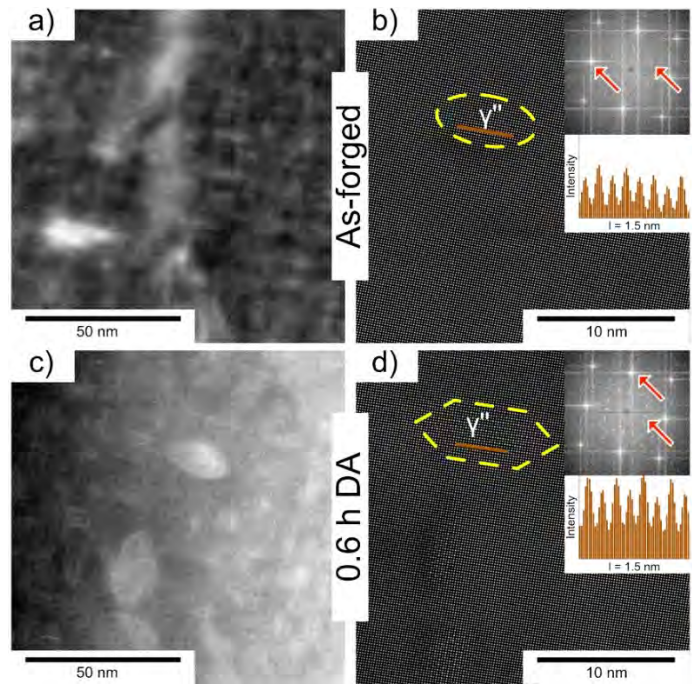

*Figure 5 STEM HAADF images of the a-b) as-forged, and c-d) 0.6 h DA conditions. Yellow outlines mark γ″-precipitates, brown lines locate the intensity plots, and red arrows mark FCC and superlattice spots in the FFT image.*

EDXS maps of the as-forged, and 0.6 h DA conditions are provided in Figure 6. The maps for Ni, Al, and Nb are shown in red, green, and blue, respectively. In the as-forged condition (Figures 6 a-d), fine and coarse precipitates exhibit particle sizes of r ~3 ± 2 nm, and ~8 ± 3 nm, respectively. Arrows indicate coarse precipitates as Ni and Nb segregation, fine precipitate structures are not resolved due to insufficient elemental partitioning. Al segregation in Figure 6 c is below the detection limit. In the 0.6 h DA condition (Figures 6 e-h), fine and coarse precipitates are identified as Ni and Nb segregation with particle sizes of r ~3 ± 1 nm, and ~10 ± 1 nm, respectively. Significant Al segregation corresponds to darker regions in the HAADF image.

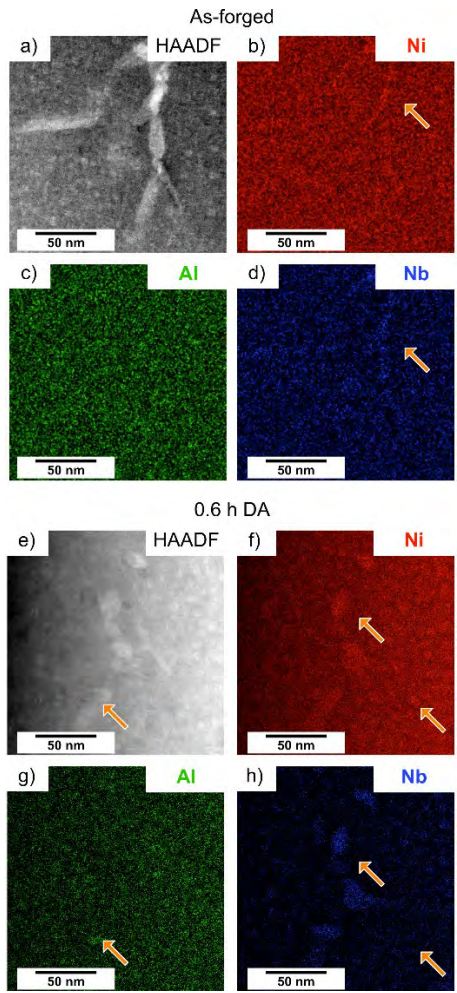

*Figure 6 HAADF and EDXS maps of the conditions a-d) as-forged, and e-h) 0.6 h DA. Arrows mark regions of segregation.*

### 3.3. Atom probe microscopy

Our voltage-mode APM data was used to generate images such as that provided in Figure 7, which are reconstructed as 2D projections of 30x30x50 nm$^3$ tomographic atom maps. Green dots represent Ni atoms in the γ-matrix, turquoise and brown iso-surfaces represent γ'- and

γ"-precipitates, respectively. The corresponding iso-surface concentration thresholds increase from the as-forged to complete DA condition, and specific values are provided in Table 2. Figures S2 a-d provide archetype 1D concentration profiles through the γ'- γ" duplet particles. One can observe the gradual separation of the γ'- and γ"-precipitates from a common γ"-nucleus.

*Table 2 Iso-surface concentration thresholds.*

| DA [h] | 0 (as-forged) | 0.2 | 0.6 | 19.1 (complete DA) |
|---|---|---|---|---|
| γ [at.%] | Ni < 59.7 ± 3.6 | Ni < 60.4 ± 2.6 | Ni < 61.5 ± 2.1 | Ni < 59.0 ± 1.6 |
| γ' [at.%] | Al+Ti > 7.5 ±1.5 | Al+Ti > 9.8 ± 2.0 | Al+Ti > 9.4 ± 2.1 | Al+Ti > 10.5 ± 0.7 |
| γ" [at.%] | Nb > 10.2 ± 1.3 | Nb > 10.2 ± 2.1 | Nb > 13.6 ± 2.6 | Nb > 13.0 ± 0.4 |

The as-forged condition in Figure 7 a exhibits fine γ'- and γ"-precipitate structures predominantly in monolithic morphology, and scarcely duplet and triplet morphology. Figure 7 b exhibits precipitates in comparable morphology for the 0.2 h DA condition. However, some coarse γ"-precipitates are found in this dataset. The 0.6 h DA condition in Figure 7 c exhibits growth of the fine monolithic γ'- and γ"-precipitates increasingly into duplet and triplet morphology. The complete DA condition in Figure 7 d shows disc-shaped γ'- and γ"-precipitates predominantly in duplet and triplet morphology, analogous to TEM images in Figures 4 g-h.

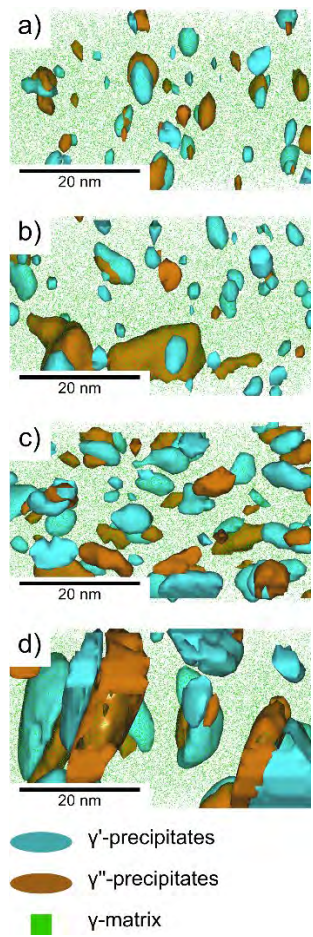

*Figure 7 Voltage-mode APM of the a) as-forged, b) 0.2 h DA, c) 0.6 h DA, and d) complete DA conditions.*

Figures 8 a-d provide large-volume datasets acquired via laser-assisted APM for representative precipitate quantification. Turquoise and brown iso-surfaces as defined by Table 2 represent γ'- and γ"-precipitates, respectively. Qualitative observations are equivalent to Figures 7 a-d, quantitative changes in the precipitate volume fractions, radii, number density and morphology are plotted in Figure 9.

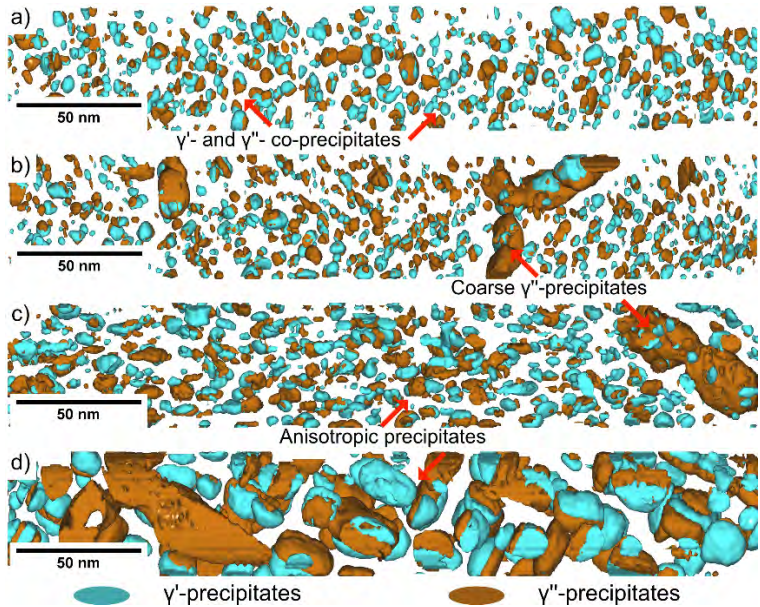

*Figure 8 Laser-mode APM reconstructions of the a) as-forged, b) 0.2 h DA, c) 0.6 h DA, and d) complete DA conditions.*

In the as-forged condition in Figure 8 a, no coarse γ''-precipitates are observed in the presented dataset. Predominantly monolithic fine γ'- and γ''-precipitates in a spheroidal-like morphology are present with volume fractions of 2.2 ± 0.7 and 1.9 ± 0.4 vol.%, respectively. The 0.2 h DA condition in Figure 8 b shows a planar array of coarse γ''-precipitates. Occasionally, coarse γ'-precipitates accompany these coarse γ''-precipitates. The surrounding volume is occupied by predominantly monolithic fine γ'- and γ''-precipitates. The volume fraction in Figure 9 a is 2.9 ± 0.5 vol.% and 4.8 ± 1.6 vol.% for γ'- and γ''-precipitates, respectively. Planar arrays of coarse γ''-precipitates exhibit further growth so that average particle radii in Figure 9 b are 3.1 ± 1.7 nm and 2.7 ± 1.7 nm for γ'- and γ''-precipitates, respectively. In Figure 9 d the complete DA condition exhibits γ'- and γ''-precipitates predominantly in duplet and triplet morphology. The particle volume fractions and radii are 13.2 ± 0.1 vol.% and 19.8 ± 4.1 vol.%, and 5.9 ± 1.2 nm and 6.6 ± 2.0 nm for γ'- and γ"-precipitates, respectively. In the course of DA, Figures 9 e-f illustrate the enrichment of Al+Ti by γ'-precipitates, and enrichment of Nb and rejection of Al+Ti by γ"-precipitates.

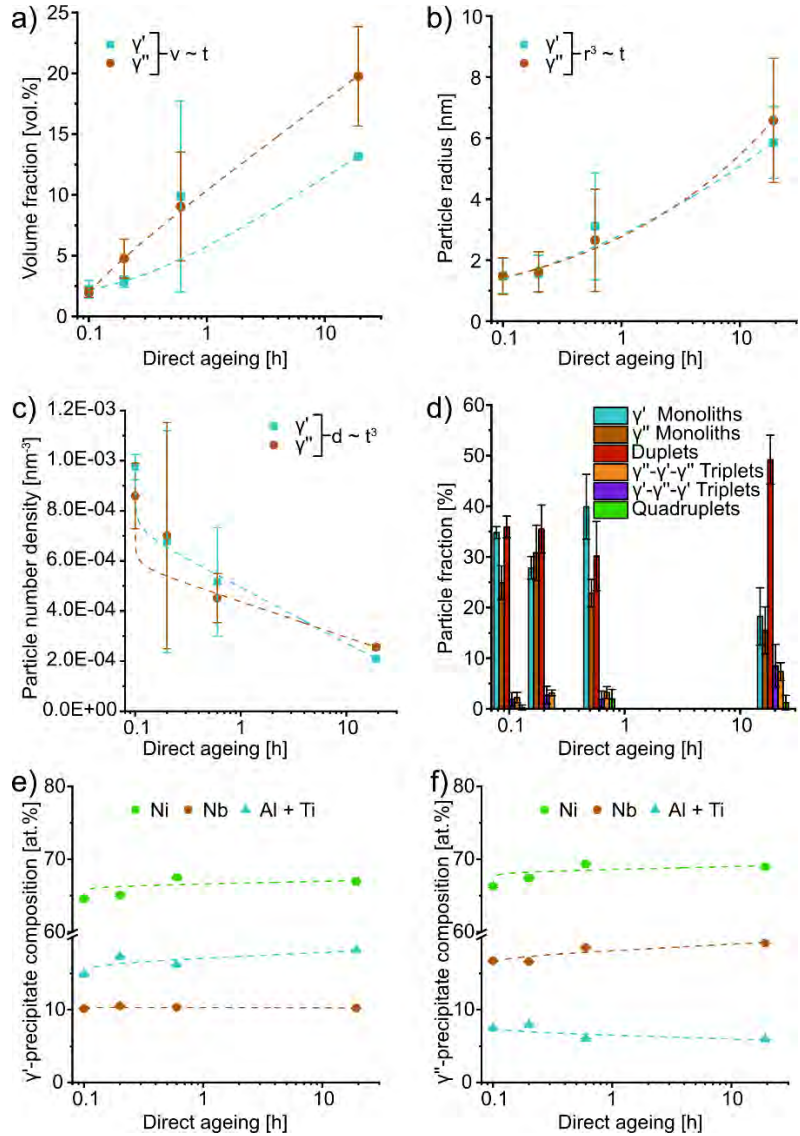

*Figure 9 Quantitative APM γ'- and γ"-precipitate statistics: a) volume fractions, b) particle radii, c) particle number density, d) morphology, and e-f) chemical compositions.*

Figure 10 provides a γ'- and γ''-precipitate morphology plot as introduced by Marceau et al. [33]. At least 40 particles were analysed for their oblateness (z/y – smallest to medium axes) and aspect ratio (y/x – medium to largest axes). The bubble radii correspond to their precipitate x-axes. The evolution from spherical to disc-shaped γ'- and γ''-precipitates is obvious. Average aspect ratio and oblateness are marked as blue (γ') and red (γ'') crosses. Detailed γ'- and γ''-precipitate radii histograms are provided in the supplementary Figure S4.

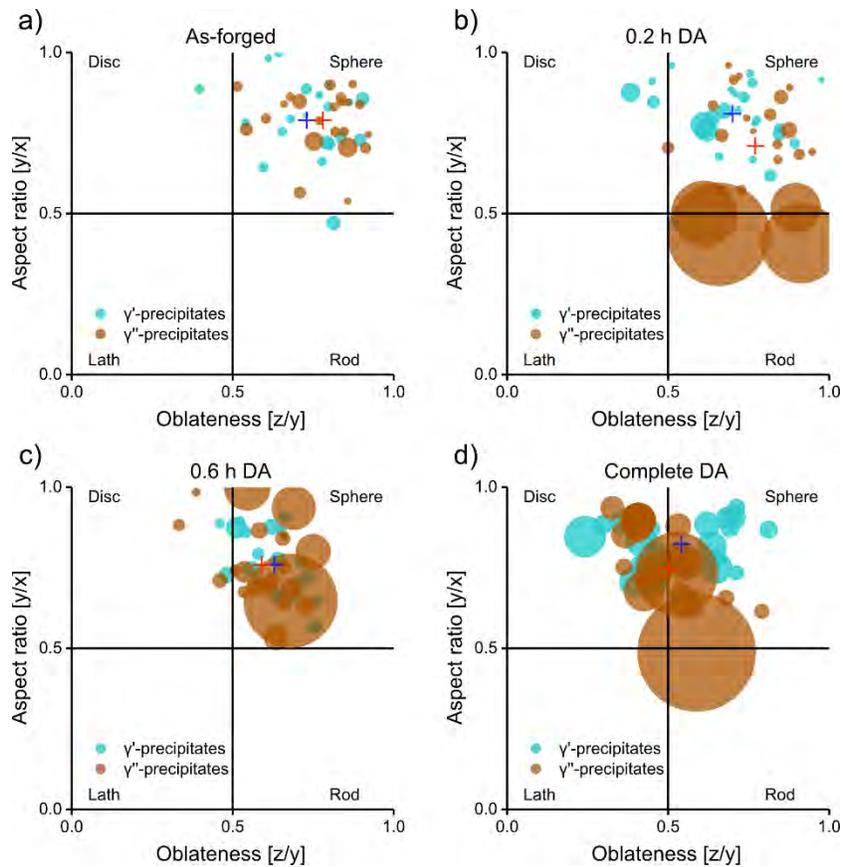

*Figure 10 Precipitate morphology for a) as-forged, b) 0.2 h DA, c) 0.6 h DA, and d) complete DA conditions. Blue and red crosses mark averages for γ' and γ'', respectively.*

The nucleation, growth and coarsening of γ'- and γ"-precipitates follows the diffusion-controlled Lifshitz-Slyozov-Wagner- (LSW) theory based on Ostwald ripening. Equation (1) has been derived for disc-shaped γ"-particles [34–36]:

$$L^3 - L_0^3 = \frac{128}{9q\pi} \frac{\Gamma C_e V_m^2 D}{k_B T} t = K" \, t \qquad (1)$$

L and $L_0$ are the mean disc diameter and the initial mean disc diameter at a given time t and Temperature T. Γ is the interfacial free energy, $C_e$ is the equilibrium solute concentration of a particle, and $V_m$ is the molar volume of γ" ($Ni_3(Nb)$). D is the diffusion coefficient, and $k_B$ is the Boltzmann constant. The aspect ratio q is defined as q = e/L. D is expressed in Equation (2).

$$D = D_0 \exp\left(\frac{-E_a}{k_B T}\right) \qquad (2)$$

$D_0$ is the diffusion factor and $E_a$ is the activation energy for particle coarsening, e.g. for γ"-precipitates. Using Equations (1) and (2), the LSW constant K" [$10^{-3}$ nm³ s⁻¹] is determined as the slope in a linear regression of $L^3$ (nm³) vs. t (s) (see supplementary Figure S3) and depends on $E_a$, Γ, $C_e$, $V_m$, and $D_0$, for a given temperature T [36].

## 4. Discussion

### 4.1. Nucleation & growth of γ'- and γ"-precipitates

The following nucleation processes of γ' and γ" are discussed in the terms of homogeneous (spontaneous within the γ-matrix), and heterogeneous (bound to features such as dislocations) nucleation. Figure 3 shows that the as-forged condition exceeds the hardness of conventionally processed Alloy 718 [22,37]. Figures 5 a-b, and Figure 8 confirm homogeneous and heterogeneous nucleation of γ' and γ" during air cooling, which has been only anticipated by Krueger [18]. In contrast, our previous study on DA of Alloy 718 [20] has shown the absence of precipitates if forging is followed by water quenching. Homogeneously nucleated particles are stable with radii of $1.3 \pm 0.6$ nm (see Figure 9 b), which are smaller than assumed by Ji et al. [38].

To the best of our knowledge, this study provides with Figures 4 and S1 novel direct evidence, that dislocations are heavily decorated by planar arrays of heterogeneously nucleated, coarse γ"-precipitates after air cooling Alloy 718 during DA. This has been observed only for conventionally processed Alloy 718 and derivatives [23,24]. Previous APM, TEM and modelling studies on Alloy 625, Alloy 706 and Alloy 718 have only assumed accelerated γ"-precipitate coarsening via Nb pipe diffusion [8,26,39]. This is an important finding, as the interaction between precipitates and retained dislocations is critical for Alloy 718 during DA [20].

Figures 9 a-c show that radii and volume fractions increase at the expense of the particle number density. The average particle radii in Figure 9 b follow the LSW-theory with $r^3 \sim t$ [40]. Average volume fraction and number density follow $v \sim t$, and $d \sim t^3$ laws, respectively. The increasingly anisotropic morphology from ~0.6 h DA shows the onset of γ'- and γ"-precipitate growth (Figures 4, 5, 7, 8). Describing γ'- and γ"-precipitates as ellipsoids in Figure 10, predominantly spherical particles (~0.8 oblateness and aspect ratio) evolve into disc-shaped particles (~0.5 oblateness and aspect ratio).

## 4.2. Coarsening γ"-precipitates

In the following, we consider populations of fine γ'-, fine γ"-, and coarse γ"-precipitates. The aspect ratios in Figure 11 a are calculated from APM data according to q in Equation (1). Fine γ'- and fine γ"-precipitates are fully coherent to the matrix and fit to earlier TEM studies [21,36,40,41]. However, coarse γ"-precipitates do not follow this trend. While the major axes L agree with the literature [21,40], the aspect ratios q are significantly increased. Larger aspect ratios have been linked to loss of coherency by more recent computational studies, and hence, are more applicable to Alloy 718 [38]. Perhaps, the apparent coalescence of coarse γ"-precipitates in earlier TEM studies has distorted the aspect ratio towards lower values. Arrays of disc-shaped particles may have been projected as one elongated particle despite their increased aspect ratios. Thus, we report loss of coherency for coarse γ"-precipitates with L ~ 38 nm .

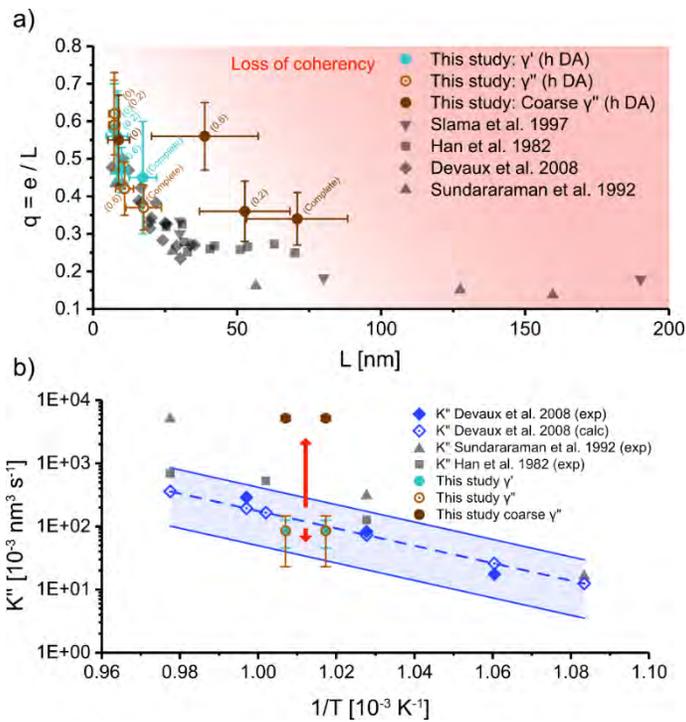

*Figure 11 a) Aspect ratios and b) K" for fine γ'- and fine γ"-precipitates and coarse γ"-precipitates* [21,36,40,41].

Figure 11 b provides K" of fine γ'- (85 ± 40 * $10^{-3}$ $nm^3$ $s^{-1}$), fine γ"- (85 ± 62 * $10^{-3}$ $nm^3$ $s^{-1}$), and coarse γ"- (5,190 ± 470 * $10^{-3}$ $nm^3$ $s^{-1}$) precipitates from 720 to 620 °C. K" of fine γ'- and fine γ"-precipitates fit below the average of previous studies [21,36,40,41]. However, K" for coarse γ"-precipitates is significantly larger than anticipated for these temperatures. Assuming $C_e$, $V_m$, $\Gamma$, and $E_a$ in Equations (1) and (2) as constants, $D_0$ must be increased by dislocations for instance via Nb pipe diffusion. In return, K" is reduced for fine γ'- and fine γ"-precipitates due to less Nb available in the γ-matrix. The present study sheds light on the influence of homogeneous versus heterogeneous nucleation on the coarsening behaviour of γ'- and γ"-precipitates during DA of Alloy 718, and thus, complements previous research [21,36,40,41].

## 4.3. Co-precipitation and morphology evolution

Figure 9 d indicates the nucleation of γ'- and γ"-precipitates predominantly in monolithic morphology, and the formation of co-precipitates in the course of DA of Alloy 718. In contrast to conventionally processed Alloy 718 [12,20], the inverted duplet γ'$_{fine}$-γ"$_{coarse}$ and triplet γ'-γ"-γ' morphologies are dominant in Figures 8 and 9.

Al+Ti are enriched by primary γ'-precipitates, and Nb is enriched and Al+Ti are rejected by primary γ"-precipitates (Figures 9 e-f). This distorts the local (Al+Ti)/Nb ratio in the proximity of the γ'/γ-matrix and γ"/γ-matrix interfaces. The high mobilities of Al and Ti in Ni have been demonstrated in diffusivity studies by Karunaratne et al. [42,43]. Alloy 718 variants with modified bulk (Al+Ti)/Nb ratios favour primary γ'-precipitates in the case of excess Al+Ti, and primary γ"-precipitates for excess Nb [15,17]. Our observations portray the local (Al+Ti)/Nb ratio as the nucleation event for secondary precipitates on the interface of the primary precipitates. Therefore, co-precipitation is a continuous process during DA of Alloy 718. Similar nucleation on γ'/γ-matrix or γ"/γ-matrix interfaces has been observed in Alloy 718 variants [17,21,44]. Consequently, we can now verify that the inverted triplet (γ'-γ"-γ') and duplet (γ'$_{fine}$-γ"$_{coarse}$) morphologies in our previous study [20] must originate from predominantly primary γ"-nucleation. Considering Figures 9 d-f and S2 a-d, γ'- and γ"-precipitates in duplet and triplet morphologies predominantly originate from a common γ"-nucleus during DA of Alloy 718. Duplet γ'$_{coarse}$-γ"$_{fine}$ and triplet γ"-γ'-γ" morphologies that have been found in conventionally aged Alloy 718 may originate from primary γ'-nucleation. However, this is beyond the scope of the current work.

## 4.4. Microstructural model

A qualitative microstructural model is derived in Figure 12. Dislocations are introduced to the γ-matrix via forging, and air-cooling results in homogeneous as well as heterogeneous γ"-nucleation events. Predominantly monolithic, fine γ'- and γ"-precipitates nucleate homogeneously in the γ-matrix, while planar arrays of coarse γ"-precipitates nucleate heterogeneously along dislocations. Due to pipe diffusion, the diffusion coefficient of Nb is significantly larger along dislocations than in the γ-matrix, so that $D_{het}(Nb) \gg D_{hom}(Nb)$. The diffusion coefficient of Al+Ti is less sensitive to dislocations, so that $D_{het}(Al+Ti) \sim D_{hom}(Al+Ti)$. Al+Ti are enriched by γ'-precipitates, and Nb is enriched and Al+Ti are rejected by γ"-precipitates. Thus, the locally changed (Al+Ti)/Nb ratio acts as nucleation event for secondary precipitates on the γ'/γ-matrix and γ"/γ-matrix interfaces. In the course of DA, fine precipitates evolve from predominantly spherical, monolithic particles into disc-shaped γ'-γ" and γ'-γ"-γ' or γ"-γ'-γ" co-precipitates with duplet and triplet morphology, respectively. Coarsening of fine γ'- and fine γ"-precipitates follows the LSW-theory, whereas growth of coarse γ"-precipitates is significantly accelerated.

The growth of coarse primary γ"-precipitates decorated by secondary γ'-precipitates is decelerated via hard impingement analogous to computational studies by Shi et al. [16]. Rapid growth of monolithic coarse γ"-precipitates results in loss of coherency to the γ-matrix.

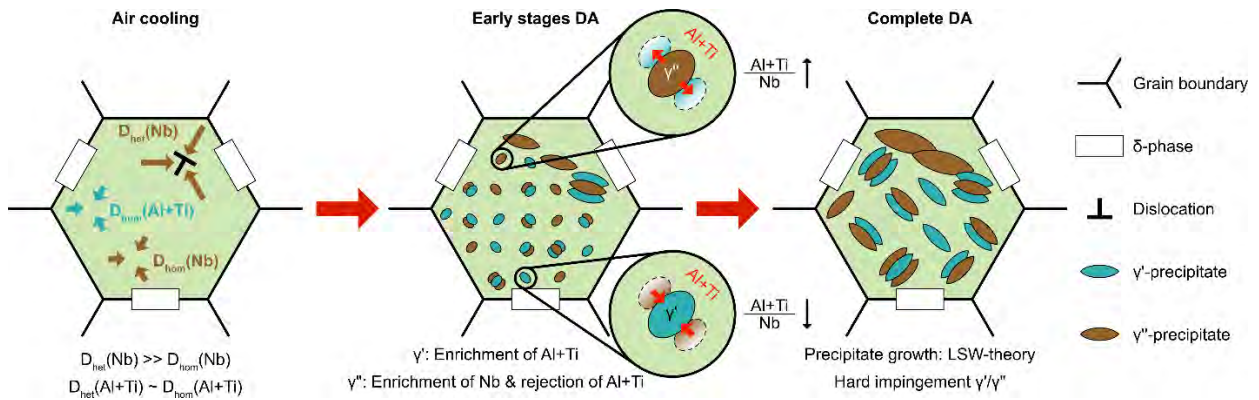

*Figure 12 Microstructural model for DA of Alloy 718 after air cooling. Pipe diffusion of Nb increases $D_{het}(Nb)$ and accelerates γ″-precipitate coarsening. Al+Ti are enriched in fine γ′-precipitates, and Nb is enriched and Al+Ti are rejected by γ″-precipitates. The distorted (Al+Ti)/Nb ratio is the nucleation event for secondary precipitates.*

## 5. Conclusions

The early stages of the direct ageing process in Alloy 718 were studied, with focus on the evolution of the γ′- and γ″-precipitation process via a correlative approach, using hardness testing, TEM, HRTEM and APM. A microstructural model is proposed to describe processes around the local changes in γ-matrix chemistry, nucleation of the primary precipitates, the role of dislocation-precipitate interactions, nucleation of the secondary precipitates, and the evolution of precipitate morphology during direct ageing. This study provides the following key findings:

(i) Fine γ′- and γ″-precipitates nucleate homogeneously as coherent spherical particles with a radius of ~1.3 nm – they remain coherent during growth into a

(i) spheroidal-like morphology, which proceeds according to the LSW-theory (K" ~85 $10^{-3}$ $nm^3$ $s^{-1}$).

(ii) Planar arrays of spheroidal-like coarse γ"-precipitates nucleate heterogeneously on dislocations and exhibit accelerated growth (K" ~5,190 $10^{-3}$ $nm^3$ $s^{-1}$) and increased aspect ratios until loss of coherency at L ~ 38 nm.

(iii) In the course of DA, Al+Ti are enriched in the γ'-precipitates, and Nb is enriched in and Al+Ti are rejected by the γ"-precipitates.

(iv) The secondary precipitates nucleate heterogeneously on the interfaces between the γ-matrix and the primary precipitates, and this is enabled by local changes in the (Al+Ti)/Nb ratio of the surrounding γ-matrix.

Our findings provide fundamental understanding about DA of Alloy 718 and offer an essential starting point for future process optimisation. Targeted micro- and nanostructure engineering may allow to suppress coarse γ"-precipitates and exploit the remaining potential of DA Alloy 718.

**Acknowledgements**

The authors acknowledge the facilities, and the scientific and technical assistance, of Microscopy Australia team (at Sydney Microscopy & Microanalysis (SMM) at The University of Sydney). The authors are especially grateful for the scientific and technical support of Drs Takanori Sato and Magnus Garbrecht (SMM) in relation to atom probe microscopy and electron microscopy, respectively. Contributions by undergraduate students

Xingying Du and Brendan Daniel (UNSW Sydney) are acknowledged. Dr Sophie Primig acknowledges research funding from the Australian Research Council via her DECRA Fellowship program under the award number DE180100440 and the UNSW Scientia Fellowship Scheme. This project has received funding from the Clean Sky 2 Joint Undertaking under the European Union's Horizon 2020 research and innovation programme under grant agreement No 714043.